# Wafer-Scale All-Dielectric quasi-BIC Metasurfaces: Bridging High-throughput Deep-UV Lithography with Nanophotonic Applications

Aidana Beisenova[1], Wihan Adi[1], Wenxin Wu[2], Shovasis K Biswas[2], Samir Rosas[1], Biljana Stamenic[3], Demis D. John[3], Filiz Yesilkoy[1,*]

[1] Department of Biomedical Engineering, University of Wisconsin-Madison, Madison, WI 53706, USA

[2] Department of Electrical and Computer Engineering, University of Wisconsin-Madison, Madison, WI 53706, USA

[3] Nanofabrication Facility, Department of Electrical and Computer Engineering, University of California, Santa Barbara, CA 93106, USA

*Corresponding author email: filiz.yesilkoy@wisc.edu

**Abstract**

High quality-factor (Q) dielectric metasurfaces operating in the visible to near-infrared range usually require sub-200 nm features, limiting their fabrication to expensive, low-throughput electron beam lithography. Here, we demonstrate wafer-scale metasurfaces fabricated using deep ultraviolet lithography (DUVL), a workhorse technology in the semiconductor industry. Using a radius and depth perturbation technique in a hole array patterned into a silicon nitride slab, we achieve quasi-bound states in the continuum (qBIC) resonances with measured Q-factors of 150. Critically, we introduce DUV exposure dose as a Q-factor engineering parameter and demonstrate how hole depth control circumvents DUVL resolution limits. Despite stochastic nanoscale variations, the fabricated metasurfaces exhibit spatial uniformity, a consequence of the nonlocal nature of the qBIC resonances. Proof of concept refractive index sensing demonstrates 129 nm/RIU sensitivity while maintaining simple CMOS camera-based resonance shift interrogation. This work bridges scalable semiconductor manufacturing with high-performance nanophotonics, establishing a practical pathway for commercializing metasurface-based sensors, on-chip spectrometers, and integrated photonic systems.

**Introduction**

Optical metasurfaces, nanoengineered thin films with subwavelength structures, have opened new possibilities to precisely control the phase[1], direction[2,3], polarization[4], and dispersion[5] properties of light. Beyond far-field wavefront modification capabilities, optical metasurfaces can be designed to confine incident light into subwavelength mode volumes, creating high quality-factor (Q) photonic cavities with long-lived resonances. Such resonant metasurfaces have become essential platforms for applications requiring enhanced light-matter interactions, including lasing[6], nonlinear frequency conversion[7], quantum photonics[8], and molecular biosensing[9–11].

There has been remarkable progress in high-Q metasurfaces utilizing diverse material combinations, including metals[12], semiconductors[13,14], dielectrics[15], and 2D materials[16]. Moreover, an expansive toolbox of geometric design strategies has accumulated, achieving various physical resonance mechanisms. Among these approaches, metasurfaces fabricated from high-refractive-index, low-loss dielectrics and leveraging bound states in the continuum (BICs) have proven particularly powerful[17,18]. While true BIC modes are non-radiative and cannot couple to free space, their radiative counterparts—quasi-BICs (qBICs)—enable tunable radiative losses[19]. QBICs can be created by introducing structural asymmetries into the constituent building blocks of the metasurfaces, providing far-field accessibility to otherwise dark modes[20], positioning them as promising candidates for next-generation photonic technologies.

Despite their compelling performance, qBIC metasurfaces face a critical scalable manufacturing barrier that has hindered them from making real-world impacts. In particular, generating resonances in the technologically significant visible to near-infrared (NIR) range requires ordered dielectric nanostructures with critical dimensions below 200 nm, a regime inaccessible to conventional ultraviolet (UV) lithography. Consequently, researchers have predominantly relied on costly and low-throughput electron beam lithography (EBL) to fabricate high-Q metasurfaces. Alternative approaches have emerged to address the scalable manufacturing bottleneck: nano imprint lithography (NIL)[21] and self-assembly processes, including glass dewetting[22] and nanosphere lithography[23,24] offer improved throughput and reduced costs. However, these methods currently lack the large-area pattern fidelity and geometric design flexibility required for complex architectures and industry-compatible continuous operation necessary for commercial viability.

Deep UV lithography (DUVL), which has been the workhorse for the semiconductor industry, presents a compelling solution to metasurface manufacturing challenges[25,26]. As a mature, high-throughput technology operating at wafer scale with sub-200nm resolution, DUVL offers scalability, reliability, and geometric versatility, well-suited for practical metasurface production. While previous reports have demonstrated DUVL-fabricated metalenses[27], beam steering devices[28], and mid-IR resonators[29], the fabrication of wafer-scale qBIC metasurfaces with resonances in the visible to near-IR range, demanding the most stringent critical dimension control at or below 200 nm, has not been achieved.

Here, we demonstrate qBIC metasurfaces operating in the visible to NIR range with sub-200 nm features, fabricated on 4-inch wafers containing multiple chips by pushing DUVL to its resolution limits (Figure 1a, b). Our approach employs a C-4 symmetry-broken "double-hole" design[30] patterned in silicon nitride ($Si_3N_4$) thin film, where alternating holes are reduced in radius by an asymmetry parameter $\Delta r$ in orthogonal lateral directions (Figure 1c). Beyond conventional radius modulation, we introduce hole depth as a complementary Q-factor tuning mechanism, achieved simply by adjusting lithographic exposure dose (Figure 1d). Shallow, partially etched holes enhance radiative coupling to free space, reducing the Q-factor of the qBIC mode (Figure 1e). This depth-tuning strategy circumvents DUVL resolution constraints that would otherwise limit further radius reduction. Remarkably, the qBIC mode exhibits robust spectral and spatial performance even with stochastic depth variations across the chip. This is a consequence of the qBIC mode's nonlocal character, where collective unit cell interactions govern the ensemble optical response. This dual-parameter control (hole radius and depth) substantially expands the accessible Q-factor design space, offering enhanced flexibility for practical deployment of DUVL-fabricated dielectric metasurfaces in real-world applications.

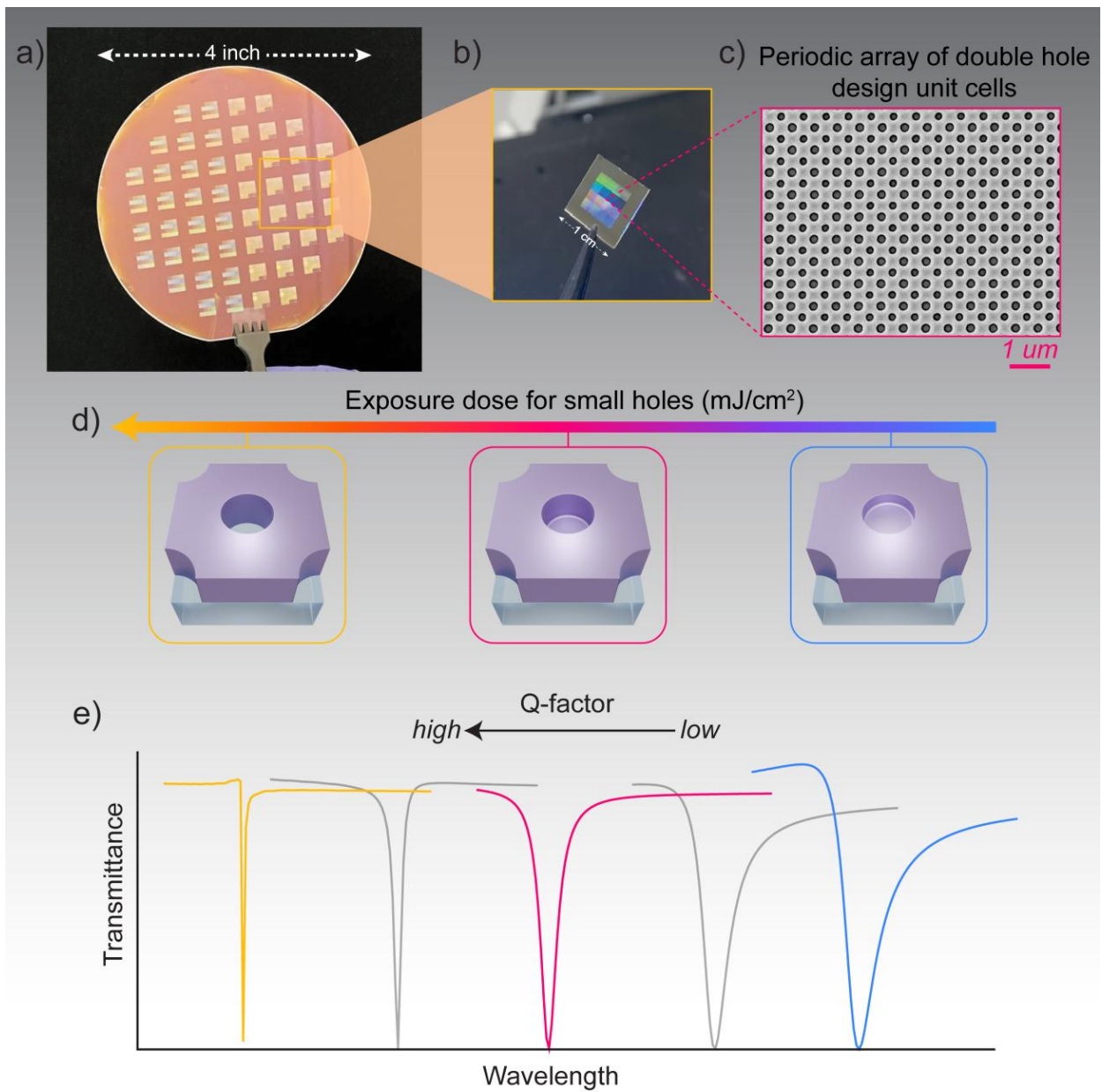

**Figure 1. Wafer-scale dielectric metasurfaces fabricated using DUVL with geometrically tunable Q-factor. a)** Photograph of a 4-inch (10 cm) wafer containing an array of dielectric metasurfaces fabricated using wafer-scale, high-throughput DUVL nanofabrication. **b)** Optical image of a diced 1 cm × 1 cm metasurface chip, illustrating the compact and scalable format of the platform. **c)** Scanning electron microscopy (SEM) image of the metasurface, showing a periodic array of double-hole unit cells etched into a $Si_3N_4$ (160 nm) slab. **d)** Schematic illustration of the effect of photolithography exposure dose on the etch depth of the smaller hole in the double-hole unit cell design. Lower doses produce shallow holes, while higher doses result in a deeper etching through the $Si_3N_4$ slab. **e)** Simulated transmission spectra demonstrating Q-factor tunability as a function of small-hole depth. Shallow holes yield broader resonances corresponding to lower Q-factors, whereas fully etched holes produce sharp resonances with high Q-factors.

First, we designed qBIC resonances based on the Brillouin zone folding (BZF) approach in a $Si_3N_4$ thin film (160 nm) deposited by low-pressure chemical vapor deposition (LPCVD) on 4-inch silica wafers to achieve high-quality, low-loss (n>1.99, k~0) dielectric material. Our metasurface design is based on periodic nanohole arrays etched into the $Si_3N_4$ film. In this reference single-hole design, a square lattice of circular holes with radius $r_1$ = 90 nm and period P = 510 nm, supports only guided-mode resonances (Figure 2a, top). Reducing the radius of every other hole by $\Delta r$ creates a double-hole design with enlarged unit cell area (P ≈ 721 nm) (Figure 2a, bottom). This symmetry breaking introduces BZF from X to X' and couples otherwise dark modes to free space, as shown in the band diagram (Figure 2c).

Simulated transmission spectra at normal incidence (Γ-point) show the emergence of sharp qBIC resonances in both TE and TM polarizations with increasing $\Delta r$ (Figure 2b). Measurements confirm the appearance of high-Q resonances under similar conditions (Figure 2d). Extracted Q-factors from simulation and experiment follow the expected inverse-square dependence, Q ≈ $a/\Delta r^2$ + b, consistent with radiative leakage induced by symmetry breaking (Figure 2e).

Near-field simulations further reveal distinct modal confinement (Figure 2f). The TM-polarized qBIC concentrates the field inside the air holes ($|E/E_0|$ ≈ 38.1), while the TE-polarized qBIC localizes near the $Si_3N_4$ surface ($|E/E_0|$ ≈ 21.3). These results establish periodic radius perturbations as a robust strategy to realize tunable high-Q resonances in all-dielectric metasurfaces.

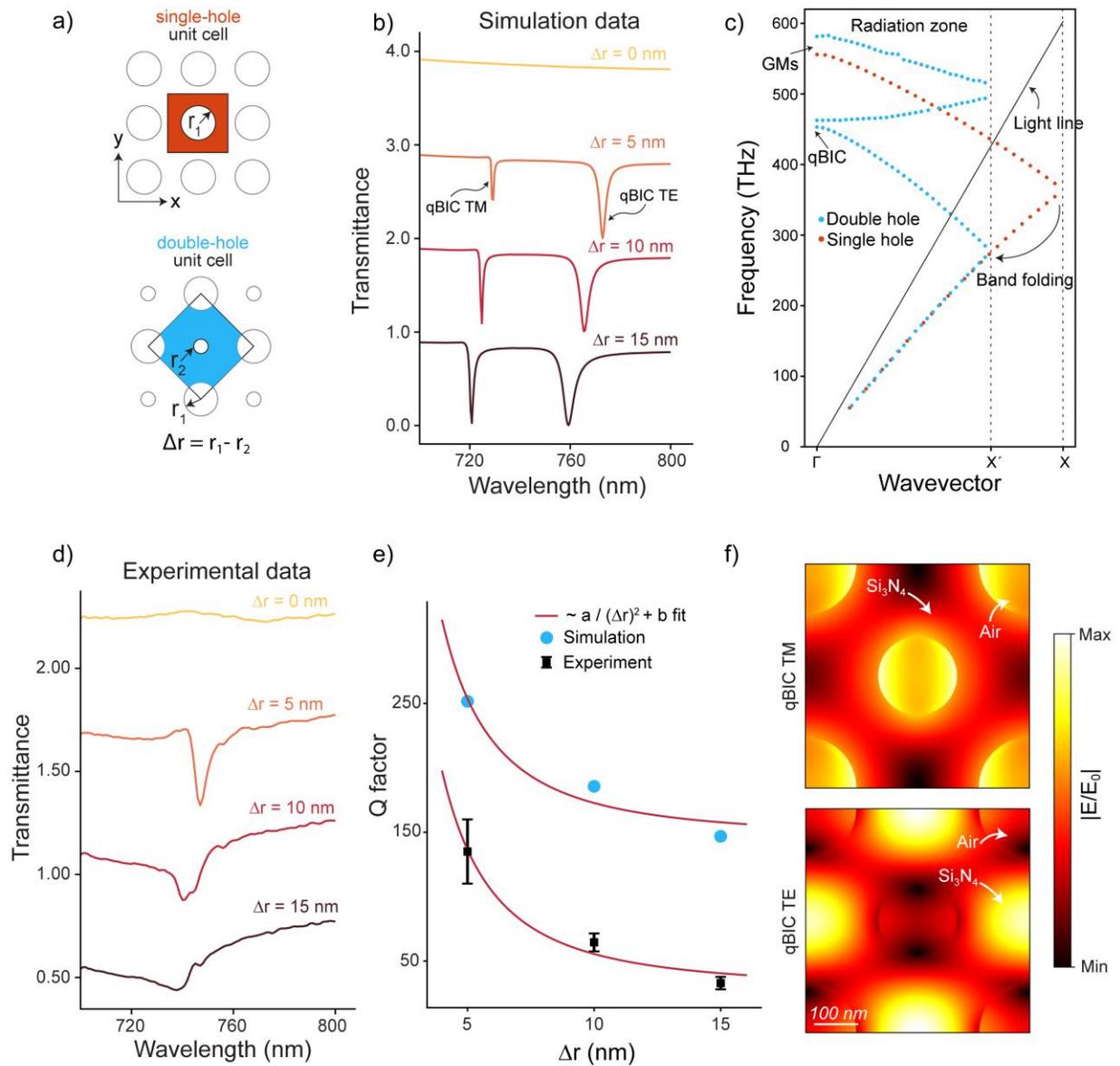

**Figure 2. Brillouin zone folding (BZF) induced quasi-bound states in the continuum (qBIC) supported by wafer-scale fabricated all-dielectric metasurfaces. a)** Schematic illustration of the square lattice designs: a symmetric single-hole unit cell and an asymmetric double-hole unit cell with radius perturbation $\Delta r = r_1 - r_2$. **b)** Simulated transmission spectra showing the emergence of qBIC modes (TM and TE polarizations) as the symmetry is broken ($\Delta r \neq 0$). **c)** Simulated band diagrams of TM modes for single-hole (blue) and double-hole (red) designs, showing band folding and the appearance of guided-mode resonances (GMRs) and qBIC modes due to symmetry breaking. **d)** Measured transmission spectra showing experimental validation of TE qBIC mode emergence with varying $\Delta r$. **e)** Q-factor dependence on asymmetry parameter $\Delta r$ extracted from both simulation (blue dots) and experiment (black markers) data. Fitted curves follow an inverse

quadratic relation: $Q \approx a / (\Delta r)^2 + b$, with fitted values a = 10789.90 $nm^2$, b = 145.70 (simulation), and a = 10789.80 $nm^2$, b = 28.50 (experiment). **f)** Simulated electric field enhancement maps ($|E/E_0|$) for the TM (top, $|E/E_0|$ = 38.10) and TE (bottom, $|E/E_0|$ = 21.32) qBIC modes, showing strong field confinement in the $Si_3N_4$-air interface. Scale bar: 100 nm.

We used a 248 nm (KrF) DUV stepper to fabricate the double-hole metasurfaces. This process requires coating the $Si_3N_4$ layer with a 60 nm bottom anti-reflection layer (BARC), followed by a 230 nm thick DUV chemically amplified photoresist. During DUV exposure, we tested different exposure doses from low (40 mJ/cm$^2$) to high (61 mJ/cm$^2$). The dose, defined as the amount of energy delivered per unit area, directly affects how well the photoresist is patterned and developed. Such "dose matrices" are a standard calibration process in nanofabrication for identifying the conditions that lead to optimal feature fidelity, sidewall profile, and etch characteristics. Subsequently, DUV resist is developed, and the BARC layer is dry-etched in an $O_2$ inductively coupled plasma (ICP) to clear the holes. Next, exposed $Si_3N_4$ in the holes was ICP etched *in situ* with a $CF_4/O_2$ and the remaining DUV resist, and BARC was stripped off.

During characterization, we observed that the exposure dose primarily influenced the small hole ($r_2$ in Figure 2), including its depth, diameter, and random occurrence per area at lower exposure doses (Figure 3a). Scanning electron microscopy (SEM) (Figure 3b) and atomic force microscopy (AFM) (Figure 3c) analyses showed that at lower doses, the small holes were often missing, or they appeared shallower than the total $Si_3N_4$ film thickness, and their occurrence was random. We explain this dose-dependent small hole depth variation and stochastic occurrence with optical proximity effects and uncontrollable interference effects resulting from the use of a coherent KrF laser source in the DUVL system (ASML PAS 5500/300). At lower doses, the small holes with radius $r_2$=90 nm receive a lower than critical dose for 100% photoresist exposure, thus after development, the resist is only partially removed from the holes. The residual resist in the small holes prevents the BARC from being fully etched in $O_2$ ICP. Consequently, this leads to varying hole depths with low-dosed small holes exhibiting shallower $Si_3N_4$ etching. Moreover, due to small spatial dose fluctuations caused by laser interference, some of the small holes do not get exposed at all, leading to the stochastic nature of small hole disappearance across the metasurface area.

As the dose increased toward 61 mJ/cm$^2$, the central holes became well defined, etched through the full 160 nm $Si_3N_4$ layer, and consistently matched the depth of the outer holes, forming a complete and uniform double-hole design. Quantitative AFM measurements of the central hole depth as a function of the exposure dose (Figure 3g) confirmed this trend, with shallow (6.9 nm on average, n=100) depths observed at 49 mJ/cm$^2$ and depths approaching 160 nm at 61 mJ/cm$^2$. This dose-dependent hole formation follows the typical contrast curve of a photoresist, where small changes in exposure energy can translate into large variations in developed resist depth, especially near the linear region of the curve.

To investigate how dose-controlled geometric variations affect the optical response, we measured transmission spectra of the fabricated double-hole metasurfaces using a tunable light source and hyperspectral imaging using a scientific CMOS camera in the 400–1000 nm spectral range. At 49 mJ/cm$^2$ exposure, the far-field spectra exhibited a pronounced transmission dip at ~760 nm, corresponding to the TE qBIC mode (Figure 3d). Counterintuitively, the resonance dip weakened with increasing dose and vanished entirely at 61 mJ/cm$^2$, contrary to expectations from a fully formed double-hole metasurface with a fixed $\Delta r$ designed to support a radiative qBIC mode.

To investigate this unexpected experimental finding, we performed numerical simulations of the double-hole unit cell ($\Delta r$ = 15 nm), varying the small hole depth according to experimental AFM measurements (Figure 3g). The simulations revealed that fully etched small holes, as achieved at 61 mJ/cm$^2$, produce significantly sharper qBIC resonances (Q-factor ~ 2637) (Figures 3e and 3f). In contrast, shallow small holes broaden the resonance and enhance radiative coupling. Although the simulations assume perfect periodicity and do not capture the stochastic hole formation observed experimentally at lower doses, the robust signal from 49 mJ/cm$^2$ demonstrates resilience to structural disorder.

We attribute the absence of detectable resonance at 61 mJ/cm$^2$ to spectral undersampling, where the qBIC mode becomes too narrow to be resolved by our optical system with a spectral resolution of 2.5 nm. At intermediate doses, such as 53 mJ/cm$^2$ and 55 mJ/cm$^2$ the resonance is partially detectable but incompletely resolved, limiting the accuracy of the measured Q-factors. These results establish that, beyond the conventional radius perturbation approach, exposure dose-controlled hole depth and occupancy provide an alternative pathway for Q-factor engineering in DUVL-manufactured metasurface platforms.

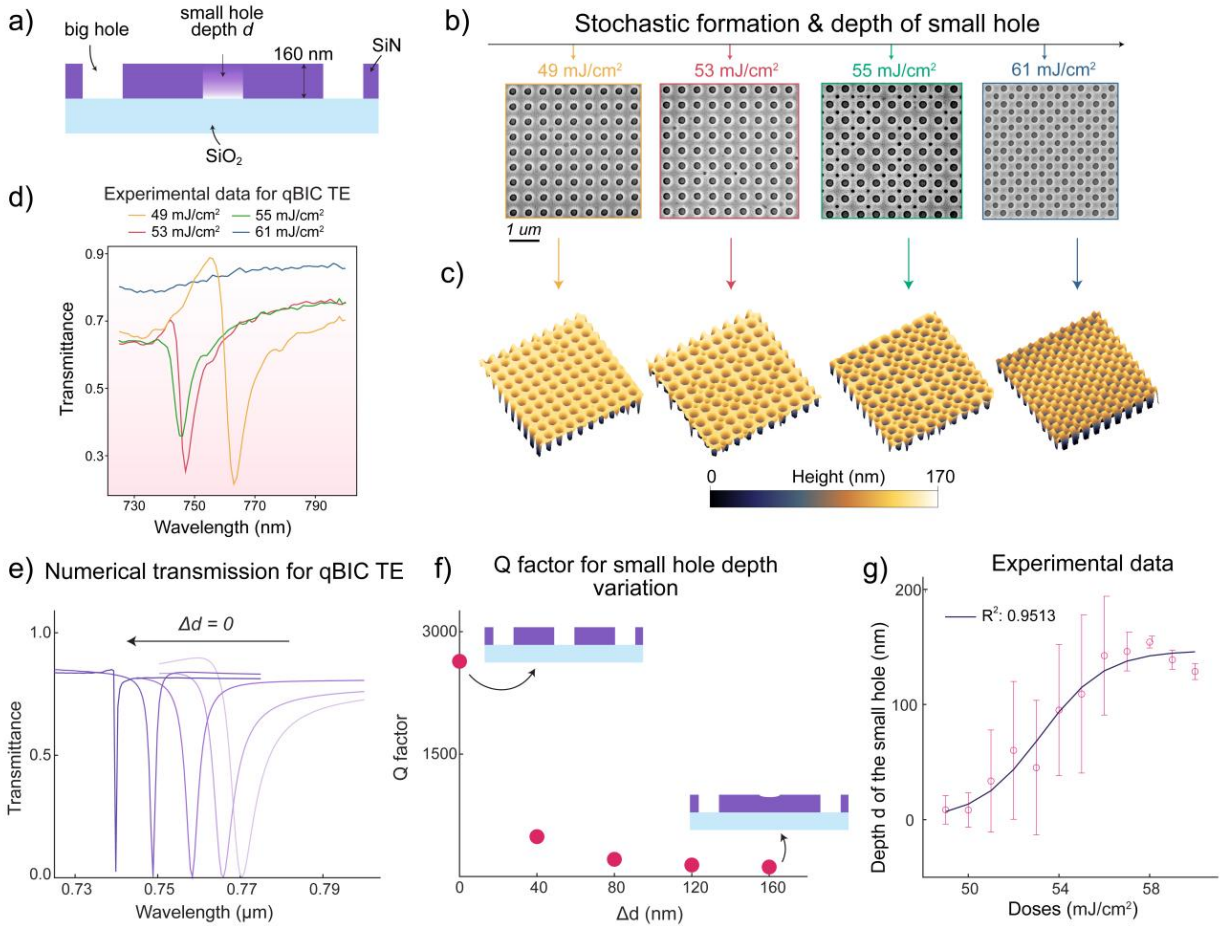

**Figure 3**. **Exposure dose-controlled Q-factor engineering of the qBIC mode in the DUVL fabricated wafer-scale dielectric metasurfaces. a)** Schematic of the simulated double-hole unit cell ($r_1$= 105 nm, $r_2$=90 nm), showing variation in small hole depth, $d$, while the big hole is kept constant in a 160 nm thick $Si_3N_4$ slab on $SiO_2$. **b)** Scanning electron microscope (SEM) images of the metasurfaces fabricated with different exposure doses (scale bar: 1 μm). At low doses (49 mJ/cm$^2$), the small hole occurrence is low and stochastic, and their depths are shallow; at higher doses (61 mJ/cm$^2$), small holes become well-defined and uniform across the array. **c)** 3D atomic force microscopy (AFM) micrographs showing increased depth and uniformity in the occurrence of the small holes from 49 mJ/cm$^2$ to 61 mJ/cm$^2$. **d)** Measured transmission spectra showing the evolution of the qBIC resonance as a function of exposure dose. The resonance becomes sharper with increasing dose, indicating a higher Q-factor, but vanishes at 61 mJ/cm$^2$ due to spectral undersampling. **e)** Simulated transmission spectra of the TE-polarized qBIC mode ($\Delta r$ = 25 nm) for different small hole depths $d$, showing increasing Q factor with increasing depth. **f)** Simulated Q-factors extracted from (e), demonstrating that the Q-factor increases with increasing small hole depth and approaches a maximum when the small hole is fully formed. **g)** Experimentally

extracted small hole depths as a function of exposure dose, obtained from AFM measurements. A fit is a sigmoidal function with $R^2$ = 0.9513.

Having identified robust resonance performance at low exposure doses where small holes are shallow and stochastically formed, we investigated whether this structural nonuniformity generates any spatial variations in resonance properties. Performing pixel-by-pixel spectral analysis, we extracted the resonance wavelength ($\lambda_{res}$) and Q-factor distributions by Fano fitting. At 49 mJ/cm² (Figure 4a, g), 53 mJ/cm² (Figure 4b, h), and 55 mJ/cm² (Figure 4c, i), both $\lambda_{res}$ and Q-factor distributions follow Gaussian distributions with tight standard deviations, indicating high spatial uniformity across the measured area (200 $\mu m^2$). The corresponding spatial maps in Figures 4d-f for $\lambda_{res}$ and Figures 4j-l for Q-factor visually confirm this consistency. The minor horizontal fringes are likely artifacts from fabrication. We excluded the 61mJ/cm² dose from this analysis because, as discussed above, the ultra-high Q factors exceed our spectral measurements' resolution limit.

This degree of spatial uniformity is striking, given the depth variations of the small hole at these doses (Figure 3g). We attribute this robustness to the nonlocal character of guided-mode resonances from which the qBIC mode is derived[31–34]. The qBIC modes emerge from coherent coupling across multiple unit cells, such that the collective optical response averages over local structural imperfections. The insensitivity to nanoscale fabrication imperfections represents a critical advantage for scalable manufacturing. The nonlocality in metasurfaces inherently relaxes lithographic tolerance requirements, enabling high-performance qBIC metasurfaces fabricated using industrial semiconductor manufacturing tools.

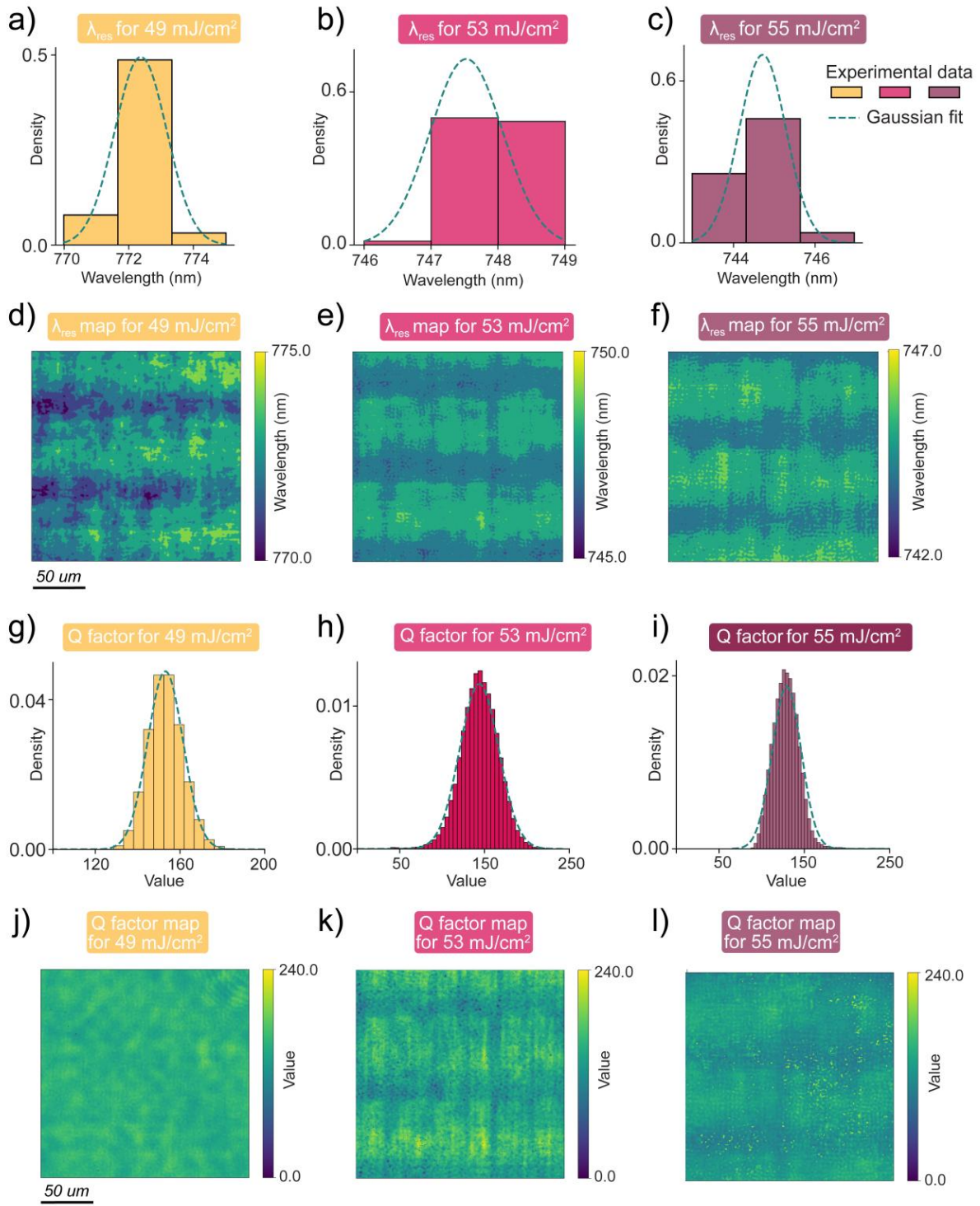

**Figure 4.** Spatial investigations of q-BIC resonance characteristics across metasurfaces fabricated with different DUV exposure doses (49 mJ/cm², 53 mJ/cm², and 55 mJ/cm²). a-c) Histograms of the resonance wavelength $\lambda_{res}$ extracted by pixel-by-pixel analysis from hyperspectral image datasets for three doses, fitted to Gaussian curves (green dashed lines). **d-f)** Spatial maps of $\lambda_{res}$ for different doses. **g-i)** Histograms of the Q-factor values corresponding to the same regions as (a–c), with Gaussian fits (green dashed lines). **j–l)** Spatial maps of the Q-factors extracted pixel-wise from hyperspectral datasets for each dose condition. The scale bar for the spatial maps is 50 μm.

Finally, to validate the refractive index sensing functionality of our platform, we performed proof-of-concept measurements by exposing the metasurface to aqueous glycerol solutions at 0-50% concentration. For each sample, we acquired hyperspectral data cubes and extracted resonance wavelengths, averaging a sensor area of 65 μm × 65 μm to ensure robust statistics. The qBIC resonance exhibits a linear redshift with increasing glycerol concentration (Figure 5a), following $\lambda_{res}$= 0.18 nm/ (% glycerol) * C + 788.36 nm, where C is the glycerol concentration (in %). This corresponds to a bulk sensitivity of 129 nm/refractive index unit (RIU), confirming the metasurface's sensitivity to refractive index changes[35].

Beyond conventional spectral tracking, we explored single-wavelength intensity readout, which is a simpler detection scheme advantageous for multiplexed or high-throughput applications. By fixing the probe wavelength at $\lambda$ = 789 nm (near the resonance inflection point) and monitoring transmittance, we observed intensity changes at different concentrations. Pixel-level intensity histograms (Figure 5b) and corresponding spatial maps (Figure 5c) show a clear trend of increasing intensity with higher glycerol content, indicating spectral red-shift through the detection window (Figure 5d). To quantitatively assess the refractive index discrimination capability of this intensity-based sensing approach, we generated receiver operating characteristic (ROC) curves from the pixel-level intensity distributions (Figure 5e). The ROC analysis demonstrates excellent classification performance, with area-under-curve (AUC) values closer to 1.00 for concentrations above 20% and 0.65 for 10%. This performance indicates that even modest refractive index changes (0.014 RIU per 10% glycerol concentrations) are reliably distinguishable at the single pixel level. Overall, these results demonstrate the wafer-scale all-dielectric metasurface's potential for label-free optical sensing using both spectral shift and single-wavelength intensity readout from hyperspectral imaging data.

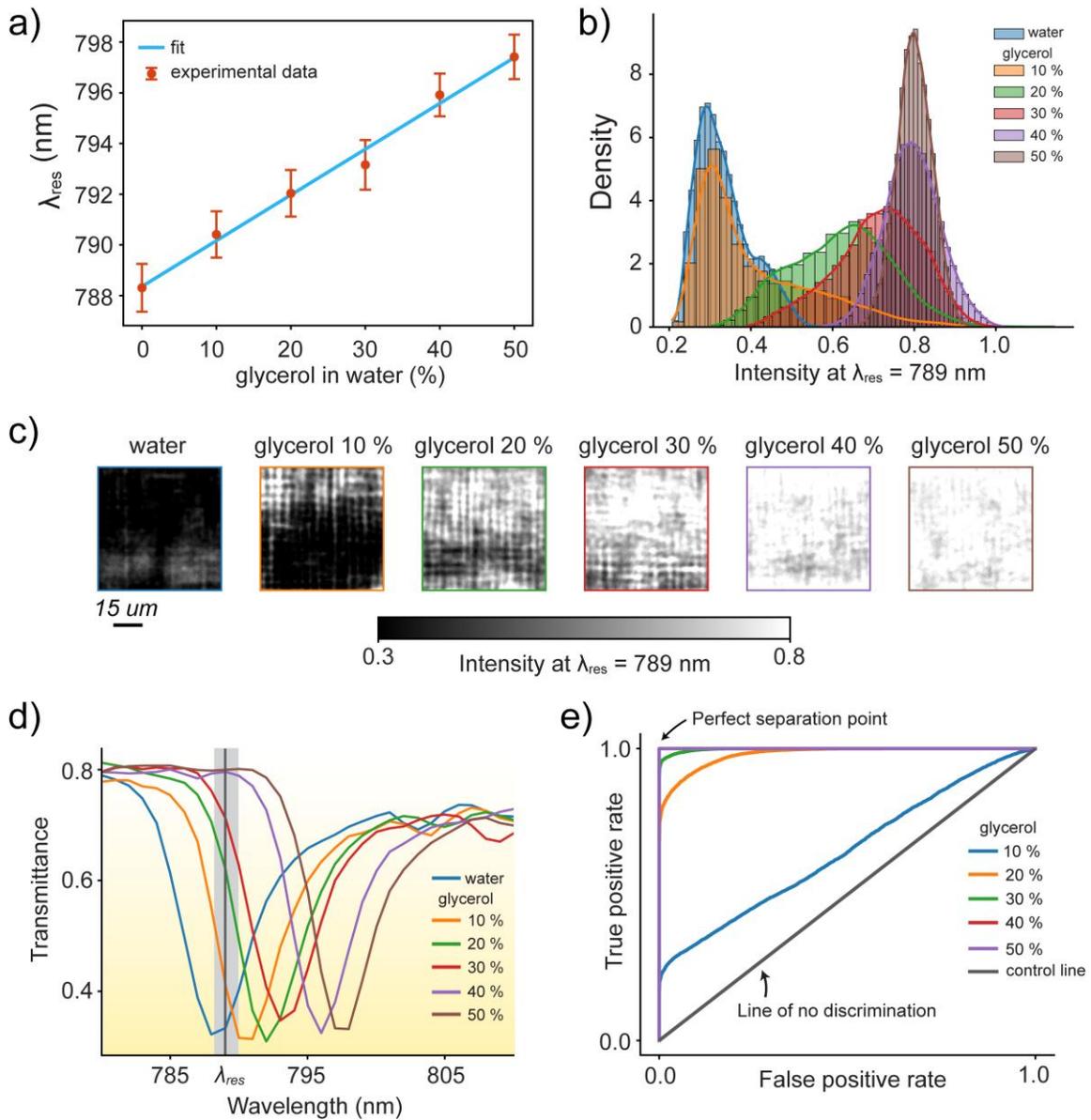

**Figure 5. Refractive index sensing with the qBIC modes of the wafer-scale fabricated metasurfaces. a)** Resonance wavelength $\lambda_{res}$ shifts linearly as a function of aqueous glycerol concentration with relation $\lambda_{res}$ = 0.18 nm/ (% glycerol) * C + 788.36 nm, where C is concentration in %. Error bars represent standard deviation extracted from pixel-level spatial measurements. **b)** Pixel-level intensity histograms at $\lambda_{res}$ = 789 nm across 65 ×65 $\mu m^2$ metasurface area for each glycerol concentration. As glycerol concentration increases, the histogram peaks shift to higher intensity values, enabling concentration discrimination. **c)** Metasurface sensor images captured at $\lambda_{res}$ = 789 nm illumination for each glycerol concentration. Scale bar: 15 μm. d) Mean transmittance spectra from 100,000 pixels for each glycerol concentration, highlighting the redshift of the qBIC resonance and the corresponding intensity variations at $\lambda_{res}$ = 789 nm. **e)**

Receiver Operating Characteristic (ROC) curves calculated from the intensity distributions in (b). The area under the ROC curve correlates with sample discrimination performance based on its glycerol concentration (refractive index), where 0.5, defined by the diagonal line, indicates no discrimination, and values close to 1.0 represent perfect classification.

In summary, we demonstrated wafer-scale fabrication of high-Q dielectric metasurfaces with less than 200 nm features, operating in the visible to near-infrared range using DUVL, a critical milestone for translating nanophotonic devices from laboratory prototypes to manufacturable technologies. By employing the Brillouin zone folding strategy, we engineered a metasurface achieving qBIC modes with measured Q-factors ~150. Our photonic device was fabricated by perforating holes in silicon nitride thin films on glass substrates, offering full compatibility with semiconductor device manufacturing infrastructure. Beyond conventional lateral geometric parameter tuning through radius perturbations, we introduced DUV lithographic exposure dose as a new control parameter for Q-factor engineering. Moreover, we demonstrated the strong tolerance of nonlocal qBIC resonances to stochastic structural imperfections across the metasurface array by investigating spatial statistics of metasurface resonance variations. Finally, our proof-of-concept refractive index sensing validates the platform's functionality as a sensor, demonstrating a sensitivity of ~129 nm/RIU and high-fidelity imaging-based sample discrimination via single-wavelength interrogation. Critically, the ability to tailor Q-factors to match detection system bandwidths enables optimized signal-to-noise ratios for specific sensing modalities, from high-resolution spectroscopy to rapid CMOS camera-based imaging.

This work removes a longstanding barrier to metasurface commercialization by bridging semiconductor-compatible manufacturing with design flexibility, previously accessible only through low-throughput techniques. The scalability, reproducibility, and cost-effectiveness of our approach position DUVL-fabricated qBIC metasurfaces as practical platforms for next-generation biosensing arrays, on-chip spectroscopy, and integrated photonic systems requiring cavity-enhanced light-matter interactions at a CMOS camera detectable wavelength regime.

**Data availability**

The data generated for this study has been provided in the main text and supplementary information. Additional information can be provided upon a reasonable request.

**Acknowledgements**

F.Y. acknowledges financial support from the U.S. National Institutes of Health (NIH grant no R35GM156309) and the Clinical and Translational Science Award (CTSA) program, through the NIH National Center for Advancing Translational Sciences (NCATS) (grant no UL1TR002373). A portion of this work was performed in the UCSB Nanofabrication Facility, an open access laboratory. The authors gratefully acknowledge the use of facilities and instrumentation in the

UW-Madison Wisconsin Center for Nanoscale Technology. The Center (wcnt.wisc.edu) is partially supported by the Wisconsin Materials Research Science and Engineering Center (NSF DMR-2309000) and the University of Wisconsin-Madison. The authors also thank Prof. Jennifer Choy and Prof. Eduardo R. Arvelo from Department of Electrical and Computer Engineering at UW-Madison, Prof. Kevin Eliceiri, and Dr. Jenu Chacko from Laboratory for Optical and Computational Instrumentation (LOCI) at UW-Madison for fruitful discussions and technical assistance.

**Author contributions**

A.B. performed numerical simulations, coordinated nanofabrication, conducted experiments, collected optical data, processed the simulation and experimental data, and drafted the paper. W.A. helped with the simulation, nanofabrication, experiments, data processing, and drafted the paper. W.W. helped with the optical characterization experiments. S.R. and S.K.B. assisted with the numerical simulations. D.D.J and B.S fabricated the metasurface. F.Y. conceived the idea, designed the experiments, supervised the project and drafted the paper. All authors revised the paper.

**Competing interest**

The authors declared no competing interest.

**Supplementary materials**

Supplementary material is available.

**References**

1. A. Arbabi et al., "Dielectric metasurfaces for complete control of phase and polarization with subwavelength spatial resolution and high transmission," Nat. Nanotechnol. **10**(11), 937–943 (2015) [doi:10.1038/nnano.2015.186].

2. N. Yu and F. Capasso, "Flat optics with designer metasurfaces," Nat. Mater. **13**(2), 139–150 (2014) [doi:10.1038/nmat3839].

3. N. Yu et al., "Light Propagation with Phase Discontinuities: Generalized Laws of Reflection and Refraction," Science **334**(6054), 333–337 (2011) [doi:10.1126/science.1210713].

4. A. H. Dorrah et al., "Metasurface optics for on-demand polarization transformations along the optical path," Nat. Photonics **15**(4), 287–296 (2021) [doi:10.1038/s41566-020-00750-2].

5. A. Fedotova et al., "Second-Harmonic Generation in Resonant Nonlinear Metasurfaces Based on Lithium Niobate," Nano Lett. **20**(12), 8608–8614 (2020) [doi:10.1021/acs.nanolett.0c03290].

6. A. Kodigala et al., "Lasing action from photonic bound states in continuum," Nature **541**(7636), 196–199 (2017) [doi:10.1038/nature20799].

7. Z. Liu et al., "High-Q Quasibound States in the Continuum for Nonlinear Metasurfaces," Phys. Rev. Lett. **123**(25), 253901 (2019) [doi:10.1103/physrevlett.123.253901].

8. A. S. Solntsev, G. S. Agarwal, and Y. S. Kivshar, "Metasurfaces for quantum photonics," Nat. Photonics **15**(5), 327–336 (2021) [doi:10.1038/s41566-021-00793-z].

9. A. Tittl et al., "Imaging-based molecular barcoding with pixelated dielectric metasurfaces," Science **360**(6393), 1105–1109 (2018) [doi:10.1126/science.aas9768].

10. F. Yesilkoy et al., "Ultrasensitive hyperspectral imaging and biodetection enabled by dielectric metasurfaces," Nat Photonics **13**(6), 390–396 (2019) [doi:10.1038/s41566-019-0394-6].

11. J. Hu et al., "Rapid genetic screening with high quality factor metasurfaces," Nat. Commun. **14**(1), 4486 (2023) [doi:10.1038/s41467-023-39721-w].

12. A. Aigner et al., "Plasmonic bound states in the continuum to tailor light-matter coupling," Sci. Adv. **8**(49), eadd4816 (2022) [doi:10.1126/sciadv.add4816].

13. S. Rosas et al., "Enhanced biochemical sensing with high- Q transmission resonances in free-standing membrane metasurfaces," Optica **12**(2), 178 (2025) [doi:10.1364/optica.549393].

14. W. Adi et al., "Trapping light in air with membrane metasurfaces for vibrational strong coupling," Nat. Commun. **15**(1), 10049 (2024) [doi:10.1038/s41467-024-54284-0].

15. C. Nicolaou et al., "Enhanced detection limit by dark mode perturbation in 2D photonic crystal slab refractive index sensors," Opt. Express **21**(25), 31698 (2013) [doi:10.1364/oe.21.031698].

16. T. Weber et al., "Intrinsic strong light-matter coupling with self-hybridized bound states in the continuum in van der Waals metasurfaces," Nat. Mater. **22**(8), 970–976 (2023) [doi:10.1038/s41563-023-01580-7].

17. K. Koshelev, A. Bogdanov, and Y. Kivshar, "Meta-optics and bound states in the continuum," Sci. Bull. **64**(12), 836–842 (2019) [doi:10.1016/j.scib.2018.12.003].

18. S. I. Azzam and A. V. Kildishev, "Photonic Bound States in the Continuum: From Basics to Applications," Adv. Opt. Mater. **9**(1) (2021) [doi:10.1002/adom.202001469].

19. C. W. Hsu et al., "Bound states in the continuum," Nat. Rev. Mater. **1**(9), 16048 (2016) [doi:10.1038/natrevmats.2016.48].

20. K. Koshelev et al., "Asymmetric Metasurfaces with High-Q Resonances Governed by Bound States in the Continuum," Phys. Rev. Lett. **121**(19), 193903 (2018) [doi:10.1103/physrevlett.121.193903].

21. D. K. Oh et al., "Nanoimprint lithography for high-throughput fabrication of metasurfaces," Front. Optoelectron. **14**(2), 229–251 (2021) [doi:10.1007/s12200-021-1121-8].

22. T. D. Gupta et al., "Self-assembly of nanostructured glass metasurfaces via templated fluid instabilities," Nat. Nanotechnol. **14**(4), 320–327 (2019) [doi:10.1038/s41565-019-0362-9].

23. P. Gao et al., "Large-Area Nanosphere Self-Assembly by a Micro-Propulsive Injection Method for High Throughput Periodic Surface Nanotexturing," Nano Lett. **15**(7), 4591–4598 (2015) [doi:10.1021/acs.nanolett.5b01202].

24. M. Go et al., "Facile Fabrication of Titanium Nitride Nanoring Broad-Band Absorbers in the Visible to Near-Infrared by Shadow Sphere Lithography," ACS Appl. Mater. Interfaces **15**(2), 3266–3273 (2023) [doi:10.1021/acsami.2c17875].

25. Y. Yang et al., "Nanofabrication for Nanophotonics," ACS Nano **19**(13), 12491–12605 (2025) [doi:10.1021/acsnano.4c10964].

26. S. Palatnick, D. John, and M. Millar-Blanchaer, "Investigating pathways for deep-UV photolithography of large-area nanopost-based metasurfaces with high feature-size contrast," J. Vac. Sci. Technol. B **42**(6), 062602 (2024) [doi:10.1116/6.0003947].

27. J.-S. Park et al., "All-Glass, Large Metalens at Visible Wavelength Using Deep-Ultraviolet Projection Lithography," Nano Lett. **19**(12), 8673–8682 (2019) [doi:10.1021/acs.nanolett.9b03333].

28. J. Tao et al., "Mass-Manufactured Beam-Steering Metasurfaces for High-Speed Full-Duplex Optical Wireless-Broadcasting Communications," Adv. Mater. **34**(6), e2106080 (2022) [doi:10.1002/adma.202106080].

29. A. Leitis et al., "Wafer-Scale Functional Metasurfaces for Mid-Infrared Photonics and Biosensing," Adv. Mater. **33**(43), 2102232 (2021) [doi:10.1002/adma.202102232].

30. W. Wang et al., "Brillouin zone folding driven bound states in the continuum," Nat. Commun. **14**(1), 2811 (2023) [doi:10.1038/s41467-023-38367-y].

31. S. G. Johnson et al., "Molding the flow of light," Comput. Sci. Eng. **3**(6), 38–47 (2001) [doi:10.1109/5992.963426].

32. R. Chai et al., "Emerging Planar Nanostructures Involving Both Local and Nonlocal Modes," ACS Photonics **10**(7), 2031–2044 (2023) [doi:10.1021/acsphotonics.2c01534].

33. J.-H. Song et al., "Non-local metasurfaces for spectrally decoupled wavefront manipulation and eye tracking," Nat. Nanotechnol. **16**(11), 1224–1230 (2021) [doi:10.1038/s41565-021-00967-4].

34. A. C. Overvig, S. C. Malek, and N. Yu, "Multifunctional Nonlocal Metasurfaces," Phys. Rev. Lett. **125**(1), 017402 (2020) [doi:10.1103/physrevlett.125.017402].

35. I. P. Rosas et al., "Catalytic Dehydration of Glycerol to Acrolein over a Catalyst of Pd/LaY Zeolite and Comparison with the Chemical Equilibrium," Catalysts **7**(3), 73 (2017) [doi:10.3390/catal7030073].